\documentclass[12pt]{article}
\usepackage{amssymb}
\usepackage{psfig}
\newcommand{\beq}{\begin{equation}}
\newcommand{\eeq}{\end{equation}}
\newcommand{\beqn}{\begin{eqnarray}}
\newcommand{\eeqn}{\end{eqnarray}}
\newcommand{\bea}[1]{\beq\begin{array}{#1}}
\newcommand{\eea}{\end{array}\eeq}

\newcommand{\dual}[1]{{}^{*}{#1}}

\newcommand{\ket}[1]{|\,#1\,\rangle}

\newcommand{\diff}{\partial}

\newcommand{\cD}{{\cal D}}

\newcommand{\AP}[3]{{\it Ann. Phys. }{\bf #1} (#2) #3}
\newcommand{\NP}[3]{{\it Nucl. Phys. }{\bf #1} (#2) #3}
\newcommand{\NPPS}[3]{{\it Nucl. Phys. Proc. Suppl. }{\bf #1} (#2) #3}
\newcommand{\PL}[3]{{\it Phys. Lett. }{\bf #1} (#2) #3}

\newcommand{\PRep}[3]{{\it Phys. Rep. }{\bf #1} (#2) #3}
\newcommand{\PR}[3]{{\it Phys. Rev. }{\bf #1} (#2) #3}

\newcommand{\MPL}[3]{{\it Mod. Phys. Lett. }{\bf #1} (#2) #3}

\newcommand{\PTPS}[3]{{\it Prog. Theor. Phys. Suppl. }{\bf #1} (#2) #3}
\newcommand{\JHEP}[3]{{\it JHEP }{\bf #1} (#2) #3}

\def\A2{\langle (A_{\mu}^a)^2_{min} \rangle}
\def\a2{ \langle (A_{\mu}^a)^2 \rangle}
\def\U1{ \langle (A_{\mu})^2_{min} \rangle}
\def\u1{ \langle A_{\mu}^2 \rangle}
\def\aA{ \langle g^2 (A^a_{\mu})^2_{min} \rangle}
\begin{document}
\date{}

\title{{\bf\Large On the Emerging Phenomenology of $\A2$.}
\vskip-40mm
\rightline{\small\rm ITEP-TH-51/00}
\vskip 40mm
}

\author{F.V.~Gubarev$^{\rm a,b}$, V.I.~Zakharov$^{\rm b}$ \\
        \\
        $^{\rm a}$ {\small\it Institute of Theoretical and  Experimental Physics, B.Cheremushkinskaya 25,}\\
                   {\small\it  Moscow, 117259, Russia}\\
        $^{\rm b}$ {\small\it Max-Planck Institut f\"ur Physik, F\"ohringer Ring 6, 80805 M\"unchen, Germany}\\
                   {\small\it }
}

\maketitle

\begin{abstract}\noindent 
We discuss phenomenology of the vacuum condensate $\A2$
in pure gauge theories, where $A_{\mu}^a$ is the gauge potential.
Both Abelian and non-Abelian cases are considered.  In case of the compact $U(1)$
the non-perturbative part of the condensate $\A2$ is saturated by monopoles. 
In the non-Abelian case, a two-component picture for the condensate is presented
according to which finite values of order $\Lambda^2_{QCD}$
are associated both with large and short distances. We obtain a lower bound on the
$\A2$ by considering its change at the phase transition. Numerically, it produces an
estimate similar to other measurements.
Possible physical manifestations of the condensate are discussed.
\end{abstract}

\section{Introduction}

Vacuum condensates are commonly used to parameterize non-perturbative effects in theories with strong
couplings. Focusing on QCD, the best known vacuum condensate seems to be the quark condensate
$\langle\bar{q}q\rangle$. Perturbatively, the quark condensate is proportional to the bare quark mass,
$m_q$ and vanishes in the chiral limit $m_q=0$. Thus, a non-vanishing quark condensate signifies spontaneous
breaking of the chiral symmetry.

There are also condensates which do not vanish perturbatively. A well known example is the gluon condensate,
$\langle\alpha_s(G_{\mu\nu}^a)^2 \rangle$ where $G_{\mu\nu}^a$ is the gluon field strength tensor \cite{svz}.
Perturbatively, $\langle\alpha_s(G_{\mu\nu}^a)^2\rangle\sim \Lambda_{UV}^4$ where $\Lambda_{UV}$ is the
ultraviolet cut off. Thus, care should be exercised to separate the non-perturbative
contribution from the trivial perturbative part. There are various ways of subtracting the
perturbative contribution. Historically, the QCD sum rules were the first example of such a subtraction.
Here, one concentrates on spectral functions $\Pi_j(Q^2)$:
\beq
\Pi_j(Q^2)~=~i\int d^4x \,e^{iqx}\, \langle 0|T\{j(x),j(0)\}|0\rangle\,,
\eeq
where $q^2\equiv -Q^2$ and $j(x)$ are local currents constructed on the quark and gluon fields.
Then using the Operator Product Expansion (OPE) one obtains  at large $Q^2$: 
\beq
\label{sr}
\Pi_j(Q^2)\approx \Pi_j(Q^2)_ {parton~model}\cdot \big(1+{a_j\over \ln Q^2}+{b_j\over Q^4}\big)\,,
\eeq
where the terms of order $(\ln Q^2)^{-1}$ represent ordinary radiative corrections while $b_j$ are
proportional to the (non-perturbative) gluon condensate, with the coefficient of proportionality depending
on the current $j$. In principle, the fitted value of $\langle\alpha_s(G_{\mu\nu}^a)^2 \rangle$ can depend
on inclusion of higher orders in perturbation theory but the condensate appears to be large
numerically and not sensitive to the subtractions. 

More recently, the gluon condensate was measured on the lattice.
Here, one either subtracts the perturbation theory explicitly \cite{burgio} or keeps contribution of
particular, presumably dominating non-perturbative fluctuations \cite{langfeld}. 
In what follows, we
will always assume that the perturbative contribution to the condensates is subtracted.
Moreover, we will propose a new method to obtain a lower bound on
the condensates by considering their drop at the phase transition, see Sect. 5.

A salient feature of (\ref{sr}) is the absence of $1/Q^2$ terms. 
An obvious candidate for $d=2$ condensate is $\a2$. However,
due to the gauge invariance the spectral functions $\Pi_j(Q^2)$ cannot depend on $\a2$
which is gauge dependent.
The situation is changing if one considers a gauge non-invariant quantity, say, the gluon propagator
$\Pi_{\mu\nu}(Q^2)$ \cite{lavelle,olezczuk}. Then the $\a2$ enters the OPE and its contribution
in the $\xi$ gauge is:
\beq
\label{lavelle}
\Pi_{\mu\nu}^{A^2}~=~-(1+\xi) {N_c\pi\alpha_s \over N_c^2-1}{\a2\over Q^4}
\big(\delta_{\mu\nu}-{Q_{\mu}Q_{\nu}\over Q^2}\big).
\eeq
Since $\a2$ is gauge dependent it is not clear, however, what kind of information, if any, would be produced by
measurements of $\a2$.

For the framework which is outlined here, the crucial observation is that the minimal value of 
$(A^a_\mu)^2$ may have physical meaning\footnote{
The content of the 
paper in Ref. \cite{stodolsky} was summarized in the review article \cite{itep}.
} \cite{stodolsky,stodolsky-1}. Indeed, consider a toy model when a plane is pierced by thin vortices carrying 
non-vanishing magnetic fluxes. Then,
\beq
\oint A_{\mu} dx^{\mu}~=~(flux)~\neq~0,
\eeq
where the integral is taken over a contour surrounding one of the vortices.
It is clear then that  $(A_{\mu}^2)_{min}$ is not zero and encodes
information on the vortices. The idea that there exists a connection between
$\A2$ and topological defects is the central one for the present review.
However, the example of infinitely thin vortices in the continuum $U(1)$
theory is in fact not adequate to represent topological defects since
the corresponding action is infinite. By topological defects we will 
rather understand field configurations
characterized by infinite potentials but surviving in the vacuum.

In case of the compact $U(1)$ theory the corresponding topological defects are
Dirac strings with monopoles at the end points \cite{polyakov}.
The close connection between the $\langle(A^2_{\mu})_{min}\rangle$ and the topological defects was confirmed
in this case through lattice simulations \cite{stodolsky,stodolsky-1}, for a review see
Sect.~\ref{U1}.

Because of the asymptotic freedom, the topological defects in non-Abelian theories
are associated with infinite potentials but with finite action.
Since the topological defects are marked by potentials rather than by action,
it is partly a matter of gauge fixing which topological defects are 
relevant to the vacuum state. Magnetic monopoles are apparently of special
interest because of the dual-superconductor confinement mechanism. 
The magnetic monopoles do bring in a non-zero value of $\A2$.
In Sect.~\ref{geom-mon} we discuss the connection between the monopoles 
and $\A2$ following the paper
in Ref.~\cite{gubarev}.

An outcome of this consideration is
a two-component picture for $\A2$, see Sect.~\ref{anatomy}. Namely, contributions of order
$\Lambda_{QCD}^2$ to $\A2$ are associated both with large (of order $\Lambda_{QCD}^{-1}$) and small
(of order $a$) distances, where $a$ is the lattice spacing vanishing in the continuum limit.
The soft part of $\A2$ can be measured on the lattice by using the relations 
like (\ref{lavelle}).
First measurements of this type were reported very recently 
\cite{pene} (see also \cite{parrinello}). 
By chance, the measurements of $\a2$ were performed in the Landau gauge, which is just 
the gauge which minimizes
$(A^a_{\mu})^2$. The measurements indicate that $\A2$ is large numerically. 
We discuss phenomenological implications of this result in Sect.~\ref{measure}.
In this section we present also an independent estimate of $\A2$ by measuring
its change at the deconfinement phase transition.

It is most intriguing whether the "hard part" of $\A2$  associated with 
the ultraviolet region can
enter any physical quantity. We speculate that the hard part of $\A2$ could be related
to the $1/Q^2$ corrections discussed recently within various phenomenological
frameworks (see \cite{az,narison,ss}
and references therein). Although such a relation cannot be proven, it could, in principle,
be tested in the lattice simulations, see Sect.~\ref{short-distances}.

\section{Topological defects and $\U1$ in the Abelian case.}
\label{U1}

Let us consider first photodynamics, i.e. the theory with the Lagrangian
\beq
\label{trivial}
L~=~ {1\over 4e^2}( \diff_\mu A_\nu - \diff_\nu A_\mu)^2\,.
\eeq
Since there is no interaction at all, the coupling $e^2$ is not running.
However, one can consider  different theories with various values of $e^2$. 
Although the Lagrangian looks trivial the  physics depends actually on whether one considers
non-compact or compact versions of (\ref{trivial}).
It is only the non-compact formulation which describes free photons at any value of $e^2$,
while in the compact case there is a phase transition at $e^2\sim 1$ which is due to
the monopole condensation \cite{polyakov}  (for a review see \cite{peskin}).

Physically, the difference between the compact and non-compact versions is that in the former case the
{\it Dirac strings} carry no action and are allowed in the vacuum. 
The compact formulation naturally arises within the lattice regularization, where the action
is given in terms of the contour integrals around elementary plaquettes $p$:
\beq
\label{lattice}
S ~=~ \sum_p S_p\,, \qquad 
S_p ~=~ -\frac{1}{e^2} \; \mathrm{Re}\; \exp\, i\oint_p A_{\mu}dx^{\mu} \,.
\eeq
The standard continuum action (\ref{trivial}) emerges only in the limit of small gauge potentials 
after applying the Stokes theorem. However, the configurations for which 
\beq
\oint_p A_\mu dx^\mu ~=~ 2\pi k \,, \qquad k\in Z
\label{condition}
\eeq
for every plaquette evidently have vanishing action on the lattice.
Thus the Dirac strings which are defined by Eq.~(\ref{condition}) are  the first example of what we
call topological defects. These are field configurations which are constructed on the singular in the
continuum limit gauge potentials but which do not cost infinite action.

In fact, a closed Dirac string satisfying the condition (\ref{condition}) everywhere is not
a proper example of topological defects. The point is that such a field configuration
is to be considered as a gauge copy of the classically trivial vacuum, $A_{\mu}\equiv 0$.
In particular the closed Dirac strings would not contribute to $\U1$ since it could be gauged back
$A_{\mu}\equiv 0$ by means of the gauge transformations. Although the corresponding  gauge transformations
are singular in the continuum limit they should be admitted into the theory to match the lattice formulation, 
for more details see \cite{chernodub1}.

The physically significant topological defects are therefore Dirac strings with open ends, or monopoles.
Due to the magnetic flux conservation the action associated with the monopole is not vanishing at all but
rather diverging in the ultraviolet:
\beq
A~\sim~ {L\over e^2}\int_a^{\infty} {\bf H}^2 r^2 dr ~\sim~ {L\over e^2a}\,,
\label{suppression}
\eeq
where $L$ is the length of the monopole world-line and $a$ is the lattice spacing which plays the role of
the ultraviolet cut off.  Because of (\ref{suppression}) the monopoles are highly suppressed by the action
for $L\gg a$. However, there is an entropy factor of the order $e^{+(const)L/a}$, where 
the constant $(const)$ is of pure geometrical nature and is independent, of course, on $e^2$.
The entropy overcomes the suppression due to the action at $e^2\sim 1$ and there is a phase transition,
which is nothing else but the monopole condensation.

Moreover, the model (\ref{trivial}) provides the simplest example, where the dynamical relevance
of $\U1$ may be shown analytically. In particular, the value of $\U1$  reflects the existence of topological
defects (\ref{condition},\ref{suppression}) in the theory.

It is sufficient to consider one particular monopole trajectory. Therefore, we restrict ourselves
to the non-compact version of (\ref{trivial}). Then the monopole current $j$ may be inserted
into the vacuum via the 't~Hooft loop operator \cite{tHooft-loop}:
\beq
\label{tHooft-loop}
H(A,\Sigma_j) ~=~ \exp\{\, S(F) - S(F + 2\pi \dual{\Sigma}_j) \,\}\,, \qquad
S(F)~=~ {1\over 4 e^2} \int d^4 x ( F_{\mu\nu} )^2\,,
\eeq
where $\Sigma_j$ is a Dirac string spanned on $j$ and 
$\dual\Sigma_{\mu\nu} = \frac{1}{2} \varepsilon_{\mu\nu\lambda\rho} \Sigma_{\lambda\rho}$.
The insertion of the 't~Hooft loop is equivalent to the subtraction of the energy
of one particular Dirac string $\Sigma_j$. Next consider  the expectation value $\U1$ in the theory
\beq
\label{PF}
Z(j)~=~ \int\cD A \; H(A,\Sigma_j) \; e^{-S(F)}~=~
\int\cD A \exp\{ -{1\over 4 e^2} \int d^4 x ( F_{\mu\nu} + 2\pi \dual\Sigma_{\mu\nu})^2   \}\,.
\eeq
The gauge in which the $A_\mu^2$ is minimal is canonically fixed by the introducing the Faddeev-Popov unity:
\beq
\label{unity}
1~=~ \Delta_{FP}[\lambda , A] \; \int \cD \alpha \exp\{ -\lambda \int d^4 x (A_\mu~+~\diff_\mu\alpha)^2 \}
\eeq
into the partition function (\ref{PF}) and taking the limit $\lambda \to \infty$ afterwards.
Of course, this is equivalent to the direct fixation of the Landau gauge $\diff A = 0$.
Then the expectation value $\U1$ is to be calculated as:
\beq
\U1 ~=~ - \lim\limits_{V\to\infty}\, {1\over V}\,{\diff\over \diff m^2} \,
\left.\ln Z_V(j, m^2)\right|_{m^2=0}\,,
\eeq
where $Z_V(j, m^2)$ is a partition function, defined in the finite volume $V$:
\beq
\label{PF-1}
Z_V(j, m^2 ) ~=~
\int\cD A \;\delta(\diff A)\;
\exp\{ -{1\over 4 e^2} \int_V d^4 x \;[\;
(\diff_{[\mu} A_{\nu]}+ 2\pi \dual\Sigma_{\mu\nu})^2   + m^2 A^2_\mu
\;]\;\}\,.
\eeq
The calculation of $Z_V(j, m^2 )$ is straightforward and leads to:
\beq
\label{amin}
\U1 ~=~ \lim\limits_{V\to\infty} \, {1\over V} \, {\diff\over \diff m^2} [\;
const \cdot \ln \mathrm{det}(-\diff^2 + 4 e^2 m^2) ~+
\eeq
$$
+~ {\pi^2 \over e^2} \int_V d^4 x\; j_\mu\,{1\over -\diff^2 + 4 e^2 m^2}\,j_\mu ~+~
4\pi^2 m^2 \int_V d^4 x\; \Sigma_{\mu\nu}\,{1\over -\diff^2 + 4 e^2 m^2}\,\Sigma_{\mu\nu} \;]\,,
$$
where the first term, which is independent on the inserted monopole current, represents the perturbative
contribution to $\U1$. Note that Eq.~(\ref{amin}) contains the string dependent term,
which reflects the fact that $A^2_\mu$ is not a gauge invariant quantity. When calculated in
a particular gauge $A^2_\mu$ depends on the position of the Dirac string.
From Eq.~(\ref{amin}) we conclude that the non-perturbative part of $\U1$
\beq
\label{xi}
\zeta( e^2 ) ~=~ \U1 ~-~ \U1_{pert.}
\eeq
depends only on the dynamics of monopoles and vanishes if these topological defects are absent.
Thus $\zeta(e^2)$ should have a jump at the critical coupling although it cannot be an order
parameter of the phase transition, since the monopole density is non zero even in the Coulomb phase.

Fig.~1 represents the behavior  of the quantity $\zeta(\beta)$ as a function of the
coupling constant $\beta = 1/e^2$, calculated in the numerical simulations\footnote{
The details of the numerical calculation are not important for the present qualitative arguments.
} of the lattice compact $U(1)$ gauge model. 
It clearly demonstrates that $\U1$ is indeed serves as a proper measure of the topological defects,
at least in case of $U(1)$. A unique and nice feature of the $U(1)$ case is that the perturbative part of the
condensate can be reliably removed. Indeed, in the non-compact case
$\U1$ is entirely perturbative. Moreover, the perturbation theory is the same
in the compact and non-compact cases.

\begin{figure}[t]
\centerline{
	\psfig{file=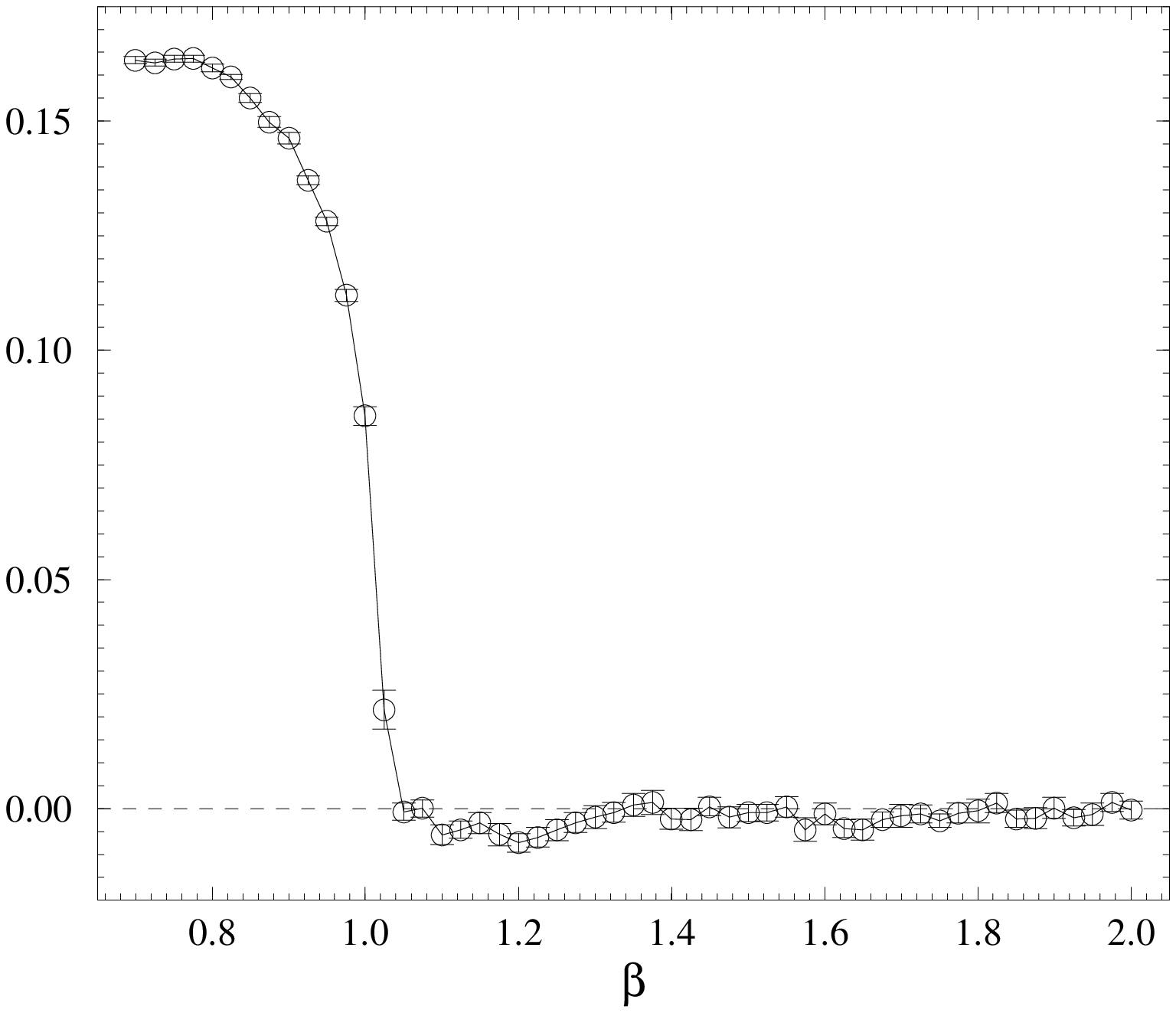,width=0.5\textwidth,silent=}
	\hspace{0.01\textwidth}
	\psfig{file=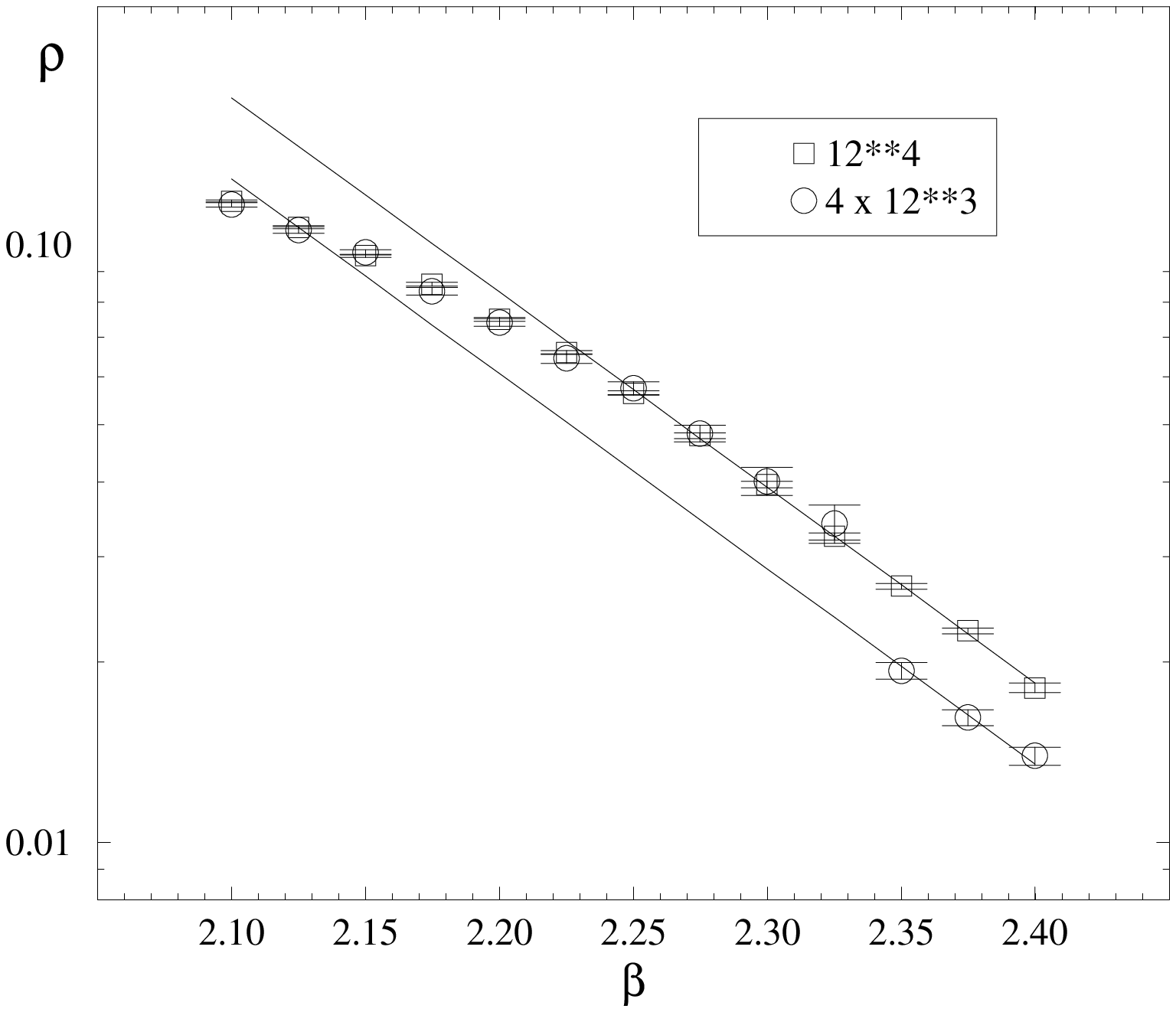,width=0.5\textwidth,silent=}
}
\noindent\fontsize{11}{13.2}\selectfont
Figure 1 (left): Nonperturbative part of $\U1$, Eq.~(\ref{xi}) in compact $U(1)$.
The phase transition occurs at $\beta = 1/e^2 \approx 1.0$.

\noindent
Figure 2 (right): The densities of geometrical monopoles in $SU(2)$ gluodynamics in the Landau gauge at zero and
finite temperatures. The solid curves are the renormalization group prediction (\ref{rg}).
\fontsize{12}{14.4}\selectfont

\end{figure}

\section{Geometrical monopoles and $\A2$}\label{geom-mon}

Turn now to the gluodynamics, 
\beq
\label{standard-SU2}
L~=~ {1\over 4g^2}(G_{\mu\nu}^a)^2\,,
\eeq
where $a=1,2,3$ is the color index (for simplicity we consider $SU(2)$ gauge group only).
If we would simply ignore the difference between the Abelian and non-Abelian cases and assume
that the non-perturbative dynamics is the same, we would run into serious difficulties.
Indeed, within the compact photodynamics $\U1\sim a^{-2}$ since the UV cut off $a$ is the only
scale in the problem. In case of the gluodynamics there exists another (hidden) scale, $\Lambda_{QCD}$,
defined in terms of the running coupling, $g^2(\Lambda_{QCD}^2)\sim 1$. Moreover, naively the Abelian-like
monopoles could not be relevant (non-perturbative) degrees of freedom in QCD since the action
(\ref{suppression}) is determined by the coupling at the UV cut off and $g^2(a^{-2})\to 0$.
As a result, the suppression of the Abelian monopoles due to their action is always stronger than the
enhancement due to the entropy factor.

The reality of the non-Abelian dynamics turns to be more varied. First, in the non-Abelian case the
Dirac strings with {\it open ends} may cost no action at all \cite{chernodub1}. The point is that although
the Abelian magnetic flux is still conserved and the Abelian part of $G_{\mu\nu}^a$ is singular at the
end points of the Dirac string, the corresponding action is not necessarily large. Indeed, there is no
direct connection any longer between the Abelian part of the field strength tensor and the action.
The commutator term can cancel and does cancel the Abelian part in the explicit construction
of Ref. \cite{chernodub1}.

Thus, in gluodynamics there are no classical monopole-like solutions similar to the Abelian Dirac
monopoles relevant to the $U(1)$ case. This does not mean, however, that the monopoles are irrelevant
in the non-Abelian case. To the contrary, the running of the coupling allows to scan the dynamics
at various values of $g^2$. Moreover, the same coupling governs dynamics of any $U(1)$ subgroup of the $SU(2)$.
In particular, if $g^2 \sim 1$ then the condensation of
Abelian-like monopoles is favored due to the entropy factor. And there is of course a lot of numerical evidence for
the relevance of the monopoles, for review and further references see \cite{review}.

Here, we will outline the geometrical monopoles introduced in Ref.~\cite{gubarev}.
Mathematically, the basic idea is to define the $U(1)$ subgroup relevant to the monopoles
locally, for each plaquette.
Physically, the crucial step is to relate the monopole properties to
the minimization of $(A_{\mu}^a)^2$.

We begin again with the observation
that the action of $SU(2)$ gluodynamics on the lattice
is constructed in terms of elementary Wilson loops, not the continuum field strength tensor:
\beq
\label{U_p}
U_p = \mathrm{P}\exp i\oint_p A_\mu dx^\mu =
e^{i F_p} = 
\cos\frac{|F_p|}{2} + i \, n_p^a\,\sigma^a \sin\frac{|F_p|}{2}\,,
\eeq
\beq
F_p =F^a_p \sigma^a/2\,,\qquad |F_p|=\sqrt{F_p^a \, F_p^a}\,,
\eeq
where $\sigma^a$ are the Pauli matrices and we have defined $n^a_p = F^a_p /|F_p|$. 
As a result, the field configurations
for which 
\beq
\label{cond-SU2}
|F_p| ~=~ 4\pi k\,, \qquad k\in Z
\eeq
cost no action and thus are the topological defects which we are looking for. Since the condition
(\ref{cond-SU2}) is quite analogous to the Abelian case (cf. Eq.~(\ref{condition})) we also refer to these
topological defects as Dirac strings. In a closely related language, one can say that the Wilson loop
(\ref{U_p}) defines a "natural" $U(1)$ associated with it as the group of rotations around the vector $n^a_p$.
In this way, one can define a $U(1)$ group for each plaquette. The definition of the $U(1)$ subgroup
varies from one plaquette to another, emphasizing the non-Abelian nature of the underlying theory.

Knowledge of the Wilson loop (\ref{U_p}) does not allow, however, to determine $k$, 
Eq.~(\ref{cond-SU2}). Indeed, the full plaquette matrix $U_p$ is
the same for $|F_p|=4\pi$ and $|F_p|=0$. What is needed is the decomposition of the plaquette variable $|F_p|$,
\beq
\label{decomposition}
|F_p|~=~ \varphi_1+\varphi_2+\varphi_3+\varphi_4\,,
\eeq
in terms of the phases $\varphi_i$, $i=1,...,4$, which are associated with the corresponding links. 
The decomposition (\ref{decomposition}) comes about naturally in the basis of the {\it coherent states}
(for details and further references see \cite{gubarev}).
Indeed, the Wilson loop $W(T)$ is defined as an evolution operator of the quantum mechanical system, the
state space of which carries the irreducible representation of the gauge group:
\beq
\label{W-l-1}
(\,\diff_t ~+~ i A \,) \ket{\psi}~=~ 0\,, \qquad
\ket{\psi(T)}~=~ W(T) \, \ket{\psi(0)}\,.
\eeq
Eq.~(\ref{W-l-1}) implies that the evolving vector $\ket{\psi(t)}$ is a generalized coherent
state. Moreover, for a given evolution operator $W(T)$ there always exists such a state $\ket{\psi_0}$
for which the entire evolution reduces to a phase factor:
\beq
\ket{\psi_0(T)} ~=~ W(T) \, \ket{\psi_0(0)} ~=~ e^{i\varphi(T)} \, \ket{\psi_0(0)}\,.
\eeq
In terms of the phase $\varphi(T)$ the fundamental Wilson loop is given by
$\mathrm{Tr~} W(T) = \cos\varphi(T)$.

In order to construct the decomposition (\ref{decomposition}) we are making use of 
the following property of the coherent states:
\beq
\label{cs-prop}
g \, \ket{\psi} ~=~ e^{i\varphi_g}\,\ket{\psi_g}\,, \qquad \forall g\in SU(2)\,,
\eeq
where $\ket{\psi_g}$ and $\varphi_g$ depend on both $\ket{\psi}$ and $g$.
Then the decomposition (\ref{decomposition}) emerges as follows. For a given plaquette matrix $U_p$ one finds
the eigenvector $\ket{\psi_0}$:
\beq
U_p \, \ket{\psi_0} ~=~ e^{i |F_p|/2}\, \ket{\psi_0}
\eeq
and then compute $|F_p|$ using Eq.~(\ref{cs-prop}):
\beq
U_p \, \ket{\psi_0} ~=~ U_1\,...\, U_4 \, \ket{\psi_0} ~=~ 
e^{i\varphi_4/2} \,U_1\,...\, U_3 \, \ket{\psi_4} ~=~ 
e^{i ( \varphi_1 + ... + \varphi_4) /2}\, \ket{\psi_0}
\eeq
As a result, for any given lattice fields configuration, one can distinguish between
$|F_p|=4\pi k $ and $|F_p|=0$ for every plaquette and detect the Dirac strings in this way.
The end points of the strings are then identified with monopoles.

The construction outlined above fully determines the Dirac strings and monopoles as geometrical objects.
As a mathematical construct, it certainly appears very appealing. However, from the physical point of view 
it is crucial that the monopoles constructed in this way are gauge dependent. Indeed, 
it is the decomposition
(\ref{decomposition}) that determines whether a particular plaquette is pierced by a 
Dirac string. But the phases $\phi_i$ are dependent on the link matrices and,
therefore, gauge dependent. Moreover, contrary to the Abelian case even the end points of the strings
(monopoles) are gauge dependent.

We need at this point a physically motivated  choice of the gauge. In view of our discussion,
the Landau gauge which minimizes the $ (A^a_\mu)^2$ seems to be singled out. Indeed, in the continuum
limit both the Dirac strings and monopoles correspond to singular gauge  potentials.
It is easy to imagine, therefore, that one can generate an  arbitrary number of spurious singularities
by going to arbitrary large potentials $A^a_\mu$, so to say inflated by the gauge transformations.
On the other hand, by minimizing $(A^a_\mu)^2$ one may hope to squeeze the number of the topological defects
to its minimum and these remaining objects  may be physically significant. 

It is amusing that the guess on the choice of the gauge can be checked through numerical simulations.
Indeed, if the geometrical monopoles are physical, then the lattice monopole density  $\rho_{lat}$
should satisfy the renormalization group equation:
\beq 
\rho_{lat}~=~ \frac{\rho_{phys}}{4\Lambda^3} \cdot \left[\frac{6\pi^2}{11}\beta\right]^{153/121} 
\exp\left( -{9\pi^2\over 11}\beta \right)\,,
\label{rg}
\eeq
where $\rho_{phys}$ is the physical density, $\beta = 4/g^2$ and the scale parameter
$\Lambda$ is fixed  by numerical value of string tension 
$a\sqrt{\sigma}= 0.1326$ at $\beta = 2.6 $ (see, e.g., Ref.~\cite{teper}).
The condition (\ref{rg}) is a very strong constraint 
on the $\rho_{lat}$ and there is no surprise
that if we do not fix the gauge in a particular way $\rho_{lat}$ does not satisfy (\ref{rg}). However the
geometrical monopoles defined in the Landau gauge turn to be physical objects, i.e. their density satisfies
the condition (\ref{rg}) numerically. On the Fig.~2 we plot
$\rho_{lat}$ in the Landau gauge versus $\beta$ on the symmetric $12^4$ and asymmetric
$4 \mathrm{~x~} 12^3$ lattices (the latter case corresponds to a finite physical temperature).
At zero temperature the lattice monopole density sharply follows Eq.~(\ref{rg}) with 
$\ln(\rho_{phys}/4\Lambda^3) \approx 12.2$, which is represented by the upper solid curve on the figure.
Therefore, we can estimate the density of geometrical monopoles in $SU(2)$ gluodynamics
at zero temperature (confinement phase):
\beq
\label{rho-1}
\rho_{phys}^{\mathrm{~(low~T)}} ~\approx~ \left(\,1.9 \sqrt{\sigma}\,\right)^3 ~\approx ~
\left(\, 840 \mathrm{~MeV}\,\right)^3\,,
\eeq
where the conventional value $ \sqrt{\sigma}=440 \mathrm{~MeV}$ has been used.
The monopole density in the high temperature deconfinement phase apparently scales in accord with
Eq.~(\ref{rg}), albeit with a somewhat smaller value of $\rho_{phys}$:
\beq
\label{rho-2}
\rho_{phys}^{\mathrm{~(high~T)}} ~\approx~ \left(\,1.7 \sqrt{\sigma}\,\right)^3 ~\approx ~
\left(\, 760 \mathrm{~MeV}\,\right)^3\,.
\eeq

To summarize, the monopoles belong both to field theory and statistical physics.
Indeed, monopoles are defined in field theoretical language. However, they are gauge
dependent and any particular monopole can be removed by a gauge transformation.
Therefore, the natural question is whether all of them can be removed by
an appropriate choice of gauge. The gauge which maximally suppresses the monopoles
is the Landau gauge defined by minimizing $(A^a_\mu)^2$. It turns out that precisely
in this gauge the monopoles become physical.

\section{Anatomy of $\A2$}\label{anatomy}

Establishing the connection between the $\A2$ and the monopole physics allows to understand better the
structure of the $\A2$ itself. Indeed, both monopoles and Dirac strings contribute now to $\A2$.
In this sense the $\A2$ condensate is basically different from, say, the gluon condensate.

As for the monopole contribution, it is divergent in the infrared for a single monopole,
\beq
\int (A_{\mu}^a)^2_{mon} d^4x ~\sim~ L \cdot R\,,
\eeq
where $L$ is the length of monopole world-line and $R$ is the infrared cut off. In reality, it means that 
the contribution of the monopoles to $\A2$ is of order $\Lambda_{QCD}^2$. Indeed, at distances
$\sim \Lambda_{QCD}^{-1}$ the approximation of the monopole gas is no longer valid.

For a Dirac string we have an estimate:
\beq 
\int (A_{\mu}^a)^2_{string}d^4x~\sim~\ln (\Lambda_{UV})\cdot L\cdot T\,,
\label{singular}
\eeq
where $\Lambda_{UV}\sim a^{-1}$ is the ultraviolet cut off, $L$ is the length and $T$ is the time of existence
of the string. The string is infinitely thin in the continuum limit and the contribution (\ref{singular}) comes
from singular potentials. There is a logarithmic divergence in the ultraviolet but this could be compensated if
we consider $\aA$ instead of $\A2$. Then we can go to the continuum limit and, 
in the logarithmic approximation,
there is no sign of the size of the string left. The contribution of the strings to $\A2$ is controlled by
typical values of $L,T$. The typical values of $L,T$ are not determined, however, 
by the dynamics of the strings
themselves since the Dirac strings carry no action\footnote{
Here we discuss only monopoles and Dirac strings living in the vacuum.
External monopoles can be introduced via the 't Hooft loop. The
corresponding Dirac strings have in the continuum limit an infinite
action, for further discussion see \cite{chernodub1}.
}. The Dirac strings emerge as a supplement to the monopoles, the dynamics of which is governed by
$g^2(\Lambda_{QCD}^2)\sim 1$. That is why $L,T\sim \Lambda_{QCD}^{-1}$ as far as $\A2$ is concerned.

It is worth emphasizing once more that no insight into the monopole dynamics can be gained through the
quasiclassical  approximation. In the classical approximation the Dirac string with open ends is not associated
with any action \cite{chernodub1}. One-loop corrections were also considered explicitly and do not distinguish
this field configuration from the trivial vacuum either. Presumably, this is true to any finite order in the
perturbation theory. 

So far we discussed connection of $\A2$ with monopoles. However, the notion of
$\A2$ may be more general than the monopole-related mechanism of confinement.
Indeed, turn to another mechanism, that is P-vortices (for review and further references see,
e.g., \cite{greensite}). In this case the P-vortices seem also to be constructed on the
topological defects, i.e. field configurations with singular potentials and no action at short distances.
Thus, generically we would get again a two-component picture for $\A2$. Numerical studies along these lines
would be very interesting.
    
\section{Measuring $\A2$}\label{measure}

As is mentioned above, one can use relations like (\ref{lavelle}) to determine $\a2$ from fits to the data.
That is what was proposed some time ago \cite{lavelle,olezczuk} and attempted very 
recently in the lattice simulations \cite{pene}.

There are two comments on this approach which we would like to add. First, it is only the soft part of the 
$\a2$ which can be treated consistently via OPE. While in reality we expect that $\a2$  is contributed also by
short distances, see above. Second, measurements in the Landau gauge are in fact singled out since then one
measures not mere gauge artifacts but rather $\A2$  which is physically meaningful.
It is just happened so that the first measurements of $\a2$ have been performed in the Landau gauge and can be,
therefore, interpreted in physical terms.

In more detail, measurements of the $1/Q^2$ corrections both to  two- and three-point Green functions
have been reported. 
In the latter case the measurements refer to the symmetrical point $G(Q^2,Q^2,Q^2)$.  
The results of the fits \cite{pene} are:
\beq
\label{fits}
\aA ~=~ ( 2.32(6) \, \mathrm{GeV}\,)^2\,, \qquad
\aA ~=~ ( 4.36(12) \,\mathrm{GeV}\,)^2\,,
\eeq
where the two numbers refer to the fits to 2- and 3-point Green functions, 
respectively. 

There is a discrepancy of factor about 4 between the two fits (\ref{fits}) which might be due to 
the yet-inconsistent treatment of higher orders
in perturbation theory \cite{pene}. Let us assume following Ref.~\cite{pene} 
that the account of the higher orders does not change the scale of the $1/Q^2$ corrections.
To appreciate the numbers (\ref{fits}) it is convenient to introduce a tachyonic
gluon mass \cite{narison}:
\beq
\langle A_{\mu}^a(-q)  A^b_{\nu}(q)\rangle  ~=~
{ \delta^{ab} \over q^2+m^2_g} \big(\delta_{\mu\nu}-{q_{\mu}q_{\nu}\over q^2}\big)
~\approx~
\big(1-{m^2_g\over q^2}\big) \cdot { \delta^{ab} \over q^2} \big(\delta_{\mu\nu}-{q_{\mu}q_{\nu}\over q^2}\big)\,.
\label{mass}
\eeq 
Then the fits (\ref{fits}) give $m_g^2 \approx (1 \div 4) \; \mathrm{GeV}^2$. Moreover, since $\a2$ measured in
the Landau gauge coincides with $\A2$ which is physical, this $m_g^2$ gives a measure of the non-perturbative
corrections. And it is then for the first time so that numerically large $1/Q^2$
corrections have been directly observed.
This result, in turn, is a confirmation of the 
theoretical speculations that there exists an ``intermediate'' mass scale 
which is formally of order $\Lambda_{QCD}^2$ but is numerically large, see \cite{itep,schmidt,shuryak}. 

In view of the discrepancy between the two fits (\ref{fits})
it would be desirable to get an independent estimate of the $\A2$.
In fact, a lower bound on $\A2$ can be obtained in a remarkably simple way.
Indeed, since the perturbative, or ultraviolet divergent part of $\A2$
is not sensitive to the phase transition, the drop in $\A2$ at the critical temperature
is due to a change in the non-perturbative part. Of course, the $A^2$ condensate might be
non-zero in both phases, but since it cannot be negative the difference between $\A2$ above
and below critical temperature provides a lower bound for $\A2$.

We have considered pure $SU(2)$ gauge theory on the $12^4$ and $4\mathrm{~x~}12^3$
lattices in the Landau gauge and measured directly the quantity
\beq
\label{eta-raz}
\eta ~=~ \frac{1}{4 V} \, \sum\limits_{x,\mu}(1-\frac{1}{2} \,\mathrm{Tr}\, U_\mu(x))\,,
\eeq
where $V$ denotes the lattice volume and $U_\mu(x)$ are the link matrices.
In the naive continuum limit (\ref{eta-raz}) reduces to 
\beq
\label{eta-dva}
\eta ~\to ~ \frac{a^2}{32} \cdot \frac{1}{V} \, \int d^4 x \; \A2
\eeq
and includes both perturbative and non-perturbative contributions.
The results of our measurements are summarized on the Fig.~3.
It is straightforward then to estimate the drop in the function $\eta(\beta)$  across the deconfinement phase
transition and obtain
\beq
\label{estimate}
\A2 ~\gtrsim ~ \frac{32}{a^2(\beta_c)} \Delta \eta ~=~ \frac{32\cdot 0.011}{a^2(\beta_c)}~=~ 
\left( \, 761.6 \mathrm{~MeV}\,\right)^2 \,.
\eeq
Numerically, the estimate (\ref{estimate}) is in agreement with (\ref{fits}).
Moreover, Eqs.~(\ref{estimate}), (\ref{rho-1}), (\ref{rho-2}) allow us to speculate that the $\A2$
condensate is mostly due to the monopoles.
Indeed, the monopole density changes across the phase transition by approximately
$20\%$, see Eqs.~(\ref{rho-1}-\ref{rho-2}). If we naively scale the change in $\A2$
with the change in the monopole density we get estimates of the $\A2$ itself, which is quite
close to (\ref{fits}). Since we treat the perturbative contributions very differently
from the paper in Ref.~\cite{pene} the result (\ref{estimate}) can be considered as an independent
confirmation of the high mass scale associated with\footnote{
Note that we used the normalization of gauge potentials (see Eq.~(\ref{standard-SU2})) different from
that in Ref.~\cite{pene} and hence there is no explicit $g^2$ factor in (\ref{estimate}). Moreover, the authors
of \cite{pene} considered $SU(3)$ gauge group, not $SU(2)$ as we did.
However, the string tension is equal in both cases and, therefore, numerically the estimate
(\ref{estimate}) should not differ considerably for $SU(2)$ and $SU(3)$.
} the $\A2$. 

\begin{figure}[t]
\centerline{\psfig{file=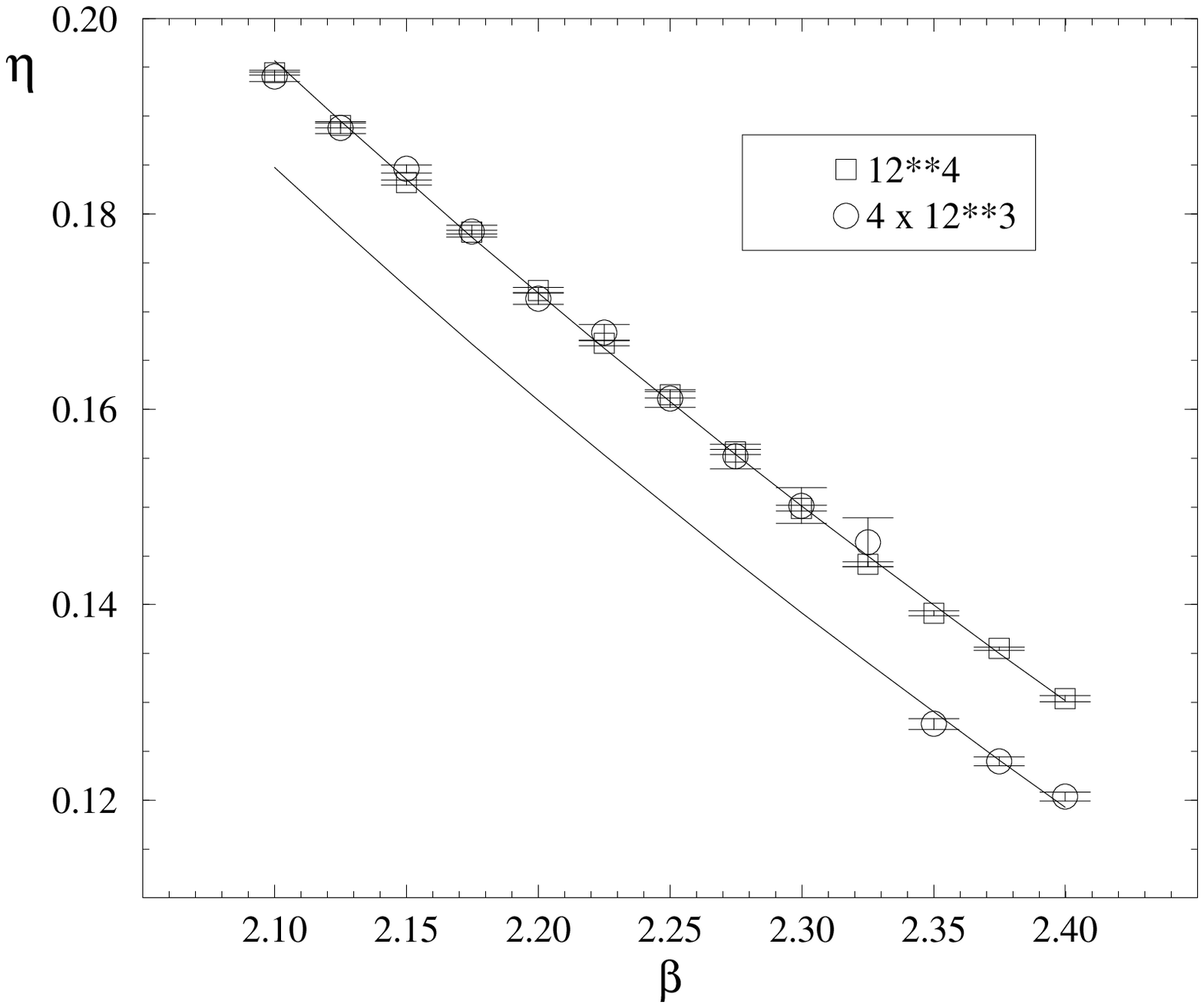,width=0.5\textwidth,silent=}}
\noindent\fontsize{11}{13.2}\selectfont
Figure 3: The quantity $\eta$, Eq.~(\ref{eta-raz}) versus $\beta$ in $SU(2)$ lattice gluodynamics.
The solid curves are drawn to guide the eye.
\fontsize{12}{14.4}\selectfont
\end{figure}

\section{Short-distance physics and $\A2$}\label{short-distances}

As is argued in Sect.~4, $\A2$ is contributed also by short distances. This contribution cannot
be treated by means of the OPE. The most intriguing question is whether this hard piece of $\A2$  may have physical
manifestations. 

The $\A2/Q^2$ corrections to the propagator can be conveniently traded for
the gluon mass, see Eq.~(\ref{mass}). In fact the notion of the gluon mass 
is more general than the Eq.~(\ref{lavelle}) based on the OPE since the $1/Q^2$
correction can be associated with short distances as well. 
Quite remarkably, the phenomenology of $1/Q^2$ corrections 
to gauge invariant quantities in terms of a tachyonic gluon mass was proposed on 
heuristic grounds \cite{narison,zakharov} and turned surprisingly successful.
Moreover, there could be a close connection between the short-distance contribution
to $\A2$ discussed above and the tachyonic mass.
Indeed evaluating the $A^2$ condensate in terms of the propagator (\ref{mass}) and subtracting the
perturbative part we get:
\beq\label{consistency}
\A2 ~\equiv~ 
\int {d^4 q\over (2\pi)^4} \; \langle A_{\mu}^a(-q) A_{\mu}^a(q) \rangle ~\sim~ m_g^2 \ln(\Lambda_{UV})
\eeq
and, therefore, $m_g^2$ conveniently parameterizes the sum of short- and large-distance contributions.
Note that the tachyonic
nature of the gluon mass reflects the positivity of the non-perturbative $\A2$ in this approach
while in Ref. \cite{narison} the tachyonic sign of the $m_g^2$ was introduced on
pure phenomenological grounds. In principle, one could hope that Eq.~(\ref{consistency})
can produce a kind of self-consistency equation. However, such an equation
would be infrared sensitive and we we would not pursue this line of investigation here.

Note that an infrared sensitive
tachyonic gluon mass was introduced first by V.N.~Gribov \cite{gribov}. The physical meaning of this mass is that
because of the hadronization gluons "decay" into hadrons. This is a pure non-perturbative effect and, within the
perturbative expansion it is described as a "leakage" from the basis  of the states used. On the theoretical side,
therefore, the central question is whether a similar interpretation is possible for the ultraviolet sensitive gluon
mass which we are discussing now. The short-distance contribution to $\A2$ comes from field configurations which
are given by singular potentials but possess no action. The Dirac strings are an example of such configurations.
The question is, therefore, whether interaction of the gluons with the Dirac strings can be characterized by a
tachyonic gluon mass.  At first sight, it is hardly possible since the gluons should not interact with  the Dirac
strings. However, if one uses the perturbative basis of the plane wave functions the gluons do interact with the
Dirac strings since the Dirac veto is not satisfied (for a recent  review and further references see \cite{chernodub2}).
In other words, accounting for the Dirac strings asks for a non-perturbative reshuffle of the basis and, within the
perturbative approach, the effect might be described by a tachyonic gluon mass. But, of course, these considerations
are highly speculative and we turn back to the phenomenology. 
 
Phenomenologically, the problem is whether one can separate, via measurements, the effect of $\A2$ entering through
ordinary OPE from the effects of the short distance tachyonic gluon mass\footnote{
In the quasiclassical approximation the non-perturbative power corrections from short distances are
due to small-size instantons and suppressed by a high power of $Q^2$, see Ref. \cite{novikov}.
This conclusion does not apply to our analysis of the monopole-related effects  since monopoles
cannot be obtained in the quasiclassical approximation, see above.
}. Physically, of course, the pictures are different since the effect of the soft part of $\A2$ cancels from the
gauge invariant quantities while the effect of short distances may well persist. However, for a non-gauge invariant 
quantity the predictions can be the same or similar.
In particular, Eq.~(\ref{mass}) was postulated in \cite{narison} and $m^2_g$ was meant to parameterize
the short-distance contribution to gauge invariant observables. Now, the propagator can be affected
also by soft part of $\A2$, see Eq.~(\ref{consistency}). Moreover, Eqs.~(\ref{mass}), (\ref{consistency})
are in fact identical. To distinguish between two contributions we will denote the short-distance part
of $m^2_g$ as $\lambda^2$.
Let us discuss the measurements of the
$1/Q^2$ corrections performed so far and their interpretation:

({\it i}) To set the scale $\lambda^2$ of the $1/Q^2$ corrections to gauge invariant quantities, let us mention
that measurements of the heavy quark potential at short distances in gluodynamics indicate:
\beq
\label{stephan}
\lambda^2 ~\sim~ 1 ~\mathrm{GeV}^2\,.
\eeq
This estimate obtained first in \cite{narison} from the data on the full potential was later confirmed by
analysis of the data on non-perturbative $\bar{Q}Q$ potential at short distances \cite{itep} (see also \cite{bali2}). 
Note that in the realistic QCD, with the effect of the light quarks included the overall fit to the correlation
functions (\ref{sr}) with inclusion of the $1/Q^2$ terms gives:
\beq
\lambda^2 ~\sim~ 0.5 ~\mathrm{GeV}^2\,.
\eeq

({\it ii}) The $1/Q^2$ corrections to the two- and three-point Green functions. 
As is already mentioned above, studying the corrections to the propagator does not
allow to separate the short- and large-distance contributions to the gluon mass.
The same is true in fact for the $G(Q^2,Q^2,Q^2)$. As is noted in Ref. \cite{pene}
the effect of the soft part of $\langle (A_{\mu}^a)^2\rangle$ on the $\alpha_s(Q^2)$ defined in terms of
the 3-point Green function at the symmetrical point is determined by OPE:
\beq
\alpha_s(Q^2)~\approx~\alpha_s^{pert}(Q^2)\cdot
\big(1-{\langle g_s^2(A_{\mu}^a)^2\rangle\over 4(N_c^2-1)}{9\over Q^2}\big)\,.
\label{three}
\eeq
Moreover the $1/Q^2$ correction in (\ref{three}) is entirely due to the renormalization of the external legs, or
renormalization factor $Z^{3/2}$. A similar equation holds for the corrections of order $\lambda^2$.
Thus, introducing $\lambda^2$ does not help to reduce the discrepancy between the two fits, see (\ref{fits}).

We pause here to note that it was not specified in Ref.~\cite{narison} in which gauge
one introduces the short distance tachyonic gluon mass. As a result, the discussion
was confined to one-gluon exchange. In view of the relation between $\langle(A_{\mu}^2)\rangle$
and the gluon mass, we would parameterize the $1/Q^2$ corrections in terms $\lambda^2$ 
specifically in the Landau gauge.

({\it iii}) The $1/Q^2$ corrections to the 3-point Green function 
at the asymmetrical point, $G(Q^2,Q^2,0)$. The corresponding measurements
are reported in \cite{parrinello} and the observed $1/Q^2$ corrections 
are numerically large. The OPE does not apply in this case and, therefore, there
are no predictions for the effect of the soft part of the $\A2$ \cite{pene}.
As for the model where the whole effect is due to a short-distance gluon mass,
the $1/Q^2$ corrections are due to the the same $Z$ factors and the same as (\ref{three}).
(The persistence of the factor $Z^{3/2}$ is due to particular definitions 
of the three-point function accepted in \cite{parrinello}). The data are indeed well
fitted by $\lambda^2\approx~ 1 \mathrm{~GeV}^2$.

To summarize our discussions, all the $1/Q^2$ corrections observed so far could be explained by
the model \cite{narison,zakharov} with a tachyonic gluon mass $\lambda^2\approx~ 1 \mathrm{~GeV}^2$. 
However, the quality of the data and their analysis is such that it is not ruled out at all
that the gauge-dependent quantities, like the gluon propagator, receive also comparable
contributions from the soft part of $\A2$ which cancels from the OPE for gauge invariant
quantities. Further measurements would hopefully clarify the situation. In particular,
measurements of the 3-point function for all external momenta large but not equal
would especially helpful.

\section*{Conclusions}

The minimal value of the potential squared, $\A2$ encodes information on 
the topological defects in gauge theories. Already first measurements 
of $\U1$ in the compact $U(1)$ \cite{stodolsky,stodolsky-1} indicate that $\A2$ 
is sensitive to field configurations responsible for the confinement.
In that case these are monopoles \cite{polyakov}, for a review see Sect. 2.

Within the dual-superconductor mechanism, monopoles play also central role
to explain the confinement in QCD.
Close connection between the $\A2$ and monopoles was revealed first in Ref.~\cite{gubarev}
in terms of the so called geometrical monopoles,
see Sect.~\ref{geom-mon}. Here, we extended the analysis of the numerical data on
the density of the geometrical monopoles at temperatures below and above
the deconfinement phase transition.
Note that in the non-Abelian case it is partly a matter of gauge fixation, which
non-perturbative fluctuations dominate in the infrared region. In particular,
analysis of possible connection between the P-vortices and $\A2$ is still awaiting
its time.

A novel feature of the $\A2$ is that even non-perturbatively it is
contributed not only by large
but small distances as well. The both contributions to $\A2$ are of order $\Lambda_{QCD}^2$.
The physics is that the density of topological defects is decided by interactions
at large distances $\sim\Lambda_{QCD}^{-1}$. If one focuses at short distances,
then the topological defects are build up on singular potentials which cost no
non-Abelian action, however. And these singular potentials bring in a finite, up to logs,
value of $\A2$.

There are various ways to measure $\A2$. First, one can attempt direct measurements
on the lattice. The main problem here is the subtraction of the trivial perturbative
contribution. The subtraction is easy to make in case of the compact $U(1)$
\cite{stodolsky,stodolsky-1},
see Sect.~\ref{U1}. In the non-Abelian case we were able to establish a lower bound
on the non-perturbative $\A2$ which is the drop in $\A2$ across the phase transition,
see Sect.~\ref{measure}.

The soft part of $\A2$ enters OPE for gauge-variant quantities like the gluon propagator
\cite{lavelle,olezczuk,pene,parrinello}. Recently, numerically large $1/Q^2$
corrections were found to two- and three-point Green functions in the Landau gauge
\cite{pene,parrinello}. Interpreted in terms of the $\A2$ the data agree
with the lower bound on $\A2$ discussed above. Moreover, the large value of the $1/Q^2$ corrections
confirms theoretical speculations that the actual scale of the violation of the
asymptotic freedom can be large numerically, see, e.g., \cite{itep,narison,ss,shuryak} and
references therein.

Separation of the short- and large-distance
contributions to $\A2$ and, more generally, to the $1/Q^2$ corrections remain a challenge
to theory. 
The main problem is to clarify how the short distance contribution to
$\A2$ enters various physical quantities. The hard part of $\A2$ could well be 
the fundamental structure behind the tachyonic mass introduced phenomenologically
in \cite{narison}. If one accepts this assumption, the short distance contribution
is sufficient numerically to explain all the existing data on the $1/Q^2$ corrections,
within existing uncertainties of the analysis.
A comparable contribution
of the soft part of the $\A2$ to gauge-variant quantities (in the Landau gauge)
is not ruled out either.

In short, we believe that already now one can conclude that the
dimension $d=2$ condensate in gauge theories, $\A2$, encodes important 
dynamical information and allows for a new
insight into the physics of both large and short distances in QCD.


\end{document}